\documentclass{article}
\usepackage{epsfig}
\usepackage{hyperref}
\catcode`\@=11
\@addtoreset{equation}{section}
\def\theequation{\thesection.\arabic{equation}}

\def\@normalsize{\@setsize\normalsize{15pt}\xiipt\@xiipt
\abovedisplayskip 14pt plus3pt minus3pt%
\belowdisplayskip \abovedisplayskip
\abovedisplayshortskip \z@ plus3pt%
\belowdisplayshortskip 7pt plus3.5pt minus0pt}

\def\small{\@setsize\small{13.6pt}\xipt\@xipt
\abovedisplayskip 13pt plus3pt minus3pt%
\belowdisplayskip \abovedisplayskip
\abovedisplayshortskip \z@ plus3pt%
\belowdisplayshortskip 7pt plus3.5pt minus0pt
\def\@listi{\parsep 4.5pt plus 2pt minus 1pt
       \itemsep \parsep
       \topsep 9pt plus 3pt minus 3pt}}

\relax
\catcode`@=12
\catcode`\@=11
\def\section{\@startsection{section}{1}{\z@}{3.5ex plus 1ex minus
     .2ex}{2.3ex plus .2ex}{\large\bf}}
\def\thesection{\arabic{section}}

\def\appendix{\setcounter{section}{0}
   \def\thesection{Appendix \Alph{section}}
   \def\theequation{\Alph{section}.\arabic{equation}}}

\begin{document}    

\begin{titlepage}
\begin{center}
{\Large  Microscopic Pictures of Dynamical Symmetry Breaking in  \\
Supersymmetric $SU(n_c)$, $USp(2n_c)$ and 
  $SO(n_c)$  Theories }
\end{center}

\vspace{1em}
\begin{center}
{\large   KENICHI KONISHI 
 }
\end{center}
\vspace{1em}
\begin{center}
{\it {
Dipartimento di Fisica, Universit\`a di Pisa   \\
  Sezione di Pisa,
        Istituto Nazionale di Fisica Nucleare, \\
Via Buonarroti, 2, Ed.B-- 56127 Pisa (Italy)\\
Department of Physics,  University of Washington, \,\,
Seattle,   WA 19185 (USA)   \\
E-mail:   konishi@phys.washington.edu;
}   }
\end{center}
\vspace{3em}
\noindent
{\bf ABSTRACT:}
{ Several distinct mechanisms of confinement and dynamical symmetry breaking (DSB) 
are identified, in a class of supersymmetric $SU(n_c)$, $USp(2n_c)$ and 
  $SO(n_c)$  gauge theories.
 In some of the vacua,
the magnetic monopoles   carrying nontrivial  flavor quantum numbers 
 condense, causing confinement and 
symmetry breaking simultaneously. In more general classes of  vacua, however, the effective low-energy
degrees of freedom are found to be constituents of the monopoles - dual (magnetic) quarks. These
magnetic quarks  condense and  give rise to confinement and  DSB.   We find two more important
 classes   of
vacua, one is in various universality classes of  nontrivial superconformal theories (SCFT), 
  another  in  free-magnetic phase.}

\vspace{1.5em}
\begin{flushleft}
Talk presented at the Workshop ``Continuous Advances in QCD"   (Univ. Minnesota)  May   2000
\end{flushleft}

\vspace{1.5em}
\begin{flushleft}
IFUP-TH/2000-19; UW/PT 00-07
\end{flushleft}

\begin{flushright}
June     2000
\end{flushright}

\end{titlepage}
\newcommand{\1}{{\Bbb I}}
\newcommand{\Z}{{\Bbb Z}}
\newcommand{\beq}{\begin{equation}}
\newcommand{\eeq}{\end{equation}}
\newcommand{\bea}{\begin{eqnarray}}
\newcommand{\eea}{\end{eqnarray}}
\newcommand{\beas}{\begin{eqnarray*}}
\newcommand{\eeas}{\end{eqnarray*}}
\newcommand{\defi}{\stackrel{\rm def}{=}}
\newcommand{\non}{\nonumber}
\def\dirac{{\cal D}}
\def\dplus{{\cal D_{+}}}
\def\dminus{{\cal D_{-}}}
\def\dbar{\bar{D}}
\def\L{{\mathcal L}}
\def\H{\cal{H}}
\def\de{\partial}
\def\si{\sigma}
\def\sb{{\bar \sigma}}
\def\rn{{\bf R}^n}
\def\r4{{\bf R}^4}
\def\s4{{\bf S}^4}
\def\ker{\hbox{\rm ker}}
\def\dim{\hbox{\rm dim}}
\def\sup{\hbox{\rm sup}}
\def\inf{\hbox{\rm inf}}
\def\infi{\infty}
\def\nrm{\parallel}
\def\nrmi{\parallel_\infty}
\def\om{\Omega}
\def\Tr{ \hbox{\rm Tr}}
\def\const{\hbox {\rm const.}}
\def\o{\over}
\def\th{\theta}
\def\im{\hbox{\rm Im}}
\def\re{\hbox{\rm Re}}
\def\bra{\langle}
\def\ket{\rangle}
\def\Arg{\hbox {\rm Arg}}
\def\Re{\hbox {\rm Re}}
\def\Im{\hbox {\rm Im}}
\def\diag{\hbox{\rm diag}}
\def\rank{\hbox {\rm Rank}}
    

\section{Introduction}

The  aim of our study here    is   to continue the search for better understanding on  the physics in the   infrared  
 of  non-Abelian  asymptotic free
gauge theories  (e.g.,  QCD). The questions of central interests   are: 
{i)} the mechanism of confinement; 
{ii)} the mechanism of flavor (chiral) symmetry breaking;
  and the relation between the two; 
{iii)} the existence of other phases (CFT, oblique confinement, etc.), 
 and   {iv)}  the $\theta$  dependence,   CP properties,  etc.
 There  has been substantial  progress recently  in  this field,
coming   from the study of supersymmetric models.\cite{SW1,SW2,DS,SUN,AD,Sei} 
 For instance,   an interesting hint came from the study of $N=2$ supersymmetric QCD with $SU(2)$
gauge group\cite{SW1,SW2}, in which supersymmetry is softly broken to $N=1$:  in some of the vacua,
condensation of   magnetic monopoles leads to  confinement and flavor   symmetry breaking
simultaneously.  

 We study here more general classes of   $N=2$ supersymmetric 
$SU(n_c)$, $USp(2n_c)$  and 
  $SO(n_c)$  gauge theories with $n_f$ quarks\cite{CKM,CKKM},      with  a small adjoint mass  breaking
supersymmetry to 
$N=1$.    The generalizations  turn out to be highly  nontrivial, and the resulting variety of dynamical
possibilities much richer  than might be expected from the $SU(2)$  cases studied by Seiberg and Witten\cite{SW1,SW2}, or 
from  the  pure ($n_f=0$)   $SU(n_c)$  theory discussed  by Douglas and Shenker\cite{DS}. 
For the exact Seiberg-Witten curves for   $N=2$ supersymmetric  $SU(n_c)$, $USp(2n_c)$  and 
  $SO(n_c)$  gauge groups see \cite{SUN}.

Before presenting the model and discuss our main results, let us mention a recent 
work   in which  the interrelation between confinement and chiral
symmetry breaking   in the standard (non-supersymmetric)   QCD  with $SU(2) $ gauge group, was
studied \cite{TaKo},  by using the    Faddeev-Niemi
gauge field decomposition.  It is  argued  that  in  the ground  state of ($SU(2)$) QCD there are two
 complementary,  competing   configurations which are important: one - meronlike configurations or regularized
Wu-Yang monopoles -  responsible  for confinement but in itself has nothing to do with
chiral symmetry breaking,   while   the instantonlike configurations   are fundamental for the chiral 
symmetry breaking but are themselves unrelated to  confinement.

\subsection{The model}\label{subsec:prod}
The models discussed here are described by the Lagrangian, 
\begin{equation}
L=  L^{(N=2)}(W,  \Phi, {\tilde Q}_i, Q^i)   +  m_i {\tilde Q}_i Q^i |_F   +
\mu  \Phi^2  |_F, 
\end{equation}
   with $m_i, \mu \ll \Lambda$, where   the first term of the standard $N=2$ supersymmetric
Lagrangian with massless hypermultiplets (quarks) in the fundamental representation of the gauge
group,   $m_i$  is the bare quark mass of the $i$-th flavor, and the adjoint mass $\mu$  breaks
supersymmetry to $N=1$.          The models have no flat directions so that  there are finite number of
isolated
$N=1$  vacua,   keeping track of which provides us with a quite nontrivial check of our analyses.   Also, 
only those theories are considered in which the interactions become strong in the infrared.  The global
symmetry of the models are:  
\begin{eqnarray}
SU(n_c): \qquad     G_F&=& U(n_f) \quad (m_i \to m, \,\,   {\hbox{\rm or}} \,\, 0);  \nonumber  \\ 
USp(2n_c): \qquad     G_F&=& SO(2 n_f) \quad (m_i \to 0);  \nonumber  \\ 
SO(n_c): \qquad   G_F&=& USp(2 n_f) \quad (m_i \to 0).
 \end{eqnarray}
Also, the global  discrete symmetries such as $Z_{2 n_c -n_f}$  in  $SU(n_c)$  play
important roles.

\subsection{The results  }\label{subsec:wpp}

 The most striking results  of our analysis,  
 summarized in Table 1  and  Table 2  for $SU(n_c)$  and  $USp(2n_c)$    theories, 
 are the  following.   
 
\begin{table}[h]
\begin{center}
\vskip .3cm
\begin{tabular}{|l|c|c|c|} 
\hline    
    Deg.Freed.      &  Eff. Gauge  Group
&   Phase    &   Global Symmetry     \\
\hline \hline
  monopoles   &   $U(1)^{n_c-1} $               &   Confinement
   &      $U(n_f) $            \\ \hline
  monopoles         & $U(1)^{n_c-1} $        &
Confinement       &     $U(n_f-1) \times  U(1) $        \\ \hline
 dual quarks        &    $SU(r)
\times U(1)^{n_c-r}   $  &    Confinement
&          $U(n_f-r) \times U(r) $
\\ \hline
  rel.  nonloc.     &    -    &    Almost SCFT
&          $U({n_f / 2} ) \times U({n_f/2}) $
\\ \hline
  dual quarks     &
$ SU({\tilde n}_c) \times  U(1)^{n_c -  {\tilde n}_c } $                &  Free Magnetic
&      $U(n_f) $         \\ \hline
\end{tabular}
\caption{  Phases of $SU(n_c)$ gauge theory with $n_f$ flavors.  The label $r$ in the third row runs 
for $r=2,3,\ldots,  [{n_f -1\over 2}].\,   $   ``rel. 
nonloc."  means that  
 relatively nonlocal monopoles and dyons   coexist  as low-energy effective degrees of freedom.   
``Confinement" and ``Free Magnetic" refer to phases with $\mu \neq 0$.
``Almost SCFT" means
that the theory is a non-trivial superconformal one for $\mu=0$ but confines with $\mu \neq 0$.  
 $ {\tilde n}_c \equiv n_f-n_c.$  }
\end{center}
\label{tabsun1}  
\end{table}     
 
\begin{table}[h]
\begin{center}
\vskip .3cm
\begin{tabular}{|l|c|c|c|}
\hline
  Deg.Freed.      &  Eff. Gauge Group
&   Phase    &   Global Symmetry     \\
\hline \hline 
  rel.  nonloc.       &    -    &
Almost SCFT
&          $ U(n_f)  $
\\ \hline
 dual quarks     &      $USp(2  {\tilde n}_c) \times  U(1)^{n_c -{\tilde n}_c} $          
     &  Free
Magnetic &      $SO(2n_f) $         \\ \hline
\end{tabular}
\caption{ Phases of $USp(2 n_c)$ gauge theory  with $n_f$ flavors  with $m_i \to 0$.  
 $ {\tilde n}_c \equiv n_f-n_c-2$. }
\label{tabuspn}
\end{center}
\end{table}

The  't Hooft - Mandelstam 
picture of confinement,  caused by     the  condensation of 
  monopoles in the maximal Abelian subgroup $U(1)^k$,  ($k=$ Rank of the gauge group),   is in fact    realized
  only in some of the vacua.  In a more
``typical" vacuum of $SU(n_c)$   gauge theory,    the  effective,  infrared degrees of freedom involve 
are   a set of dual  quarks, interacting with      low-energy  effective non-Abelian $SU(r)$   gauge
fields.  The 
condensation of these magnetic quarks as well as  of certain Abelian monopoles also present in the theory,   upon $\mu$
perturbation, lead to   confinement and dynamical symmetry breaking.  The
semi-classical monopoles may be interpreted as baryonic composites made of these magnetic quarks
and monopoles,    which  break up into
their constituents before they become massless,  as  we move from 
the semiclassical region of   the space of  $N=2$ vacua (parametrized by a set of gauge invariant VEVS)
towards the relevant singularity.

The second most interesting result is that    the special vacua  in
$SU(n_c)$ theory as well as the entire first group of vacua in $USp(2n_c)$ or   $SO(n_c)$      theory   correspond to
various   nontrivial infrared fixed points (SCFT).    The low-energy effective degrees of freedom in general contain
relatively nonlocal  states and   there is no local effective  Lagrangian description of these theories,  though the symmetry
breaking pattern can be found from the analysis at large adjoint mass $\mu$.  

 Finally, in both type of gauge  theories,   for large number
of flavors,  there is a second group of vacua in free-magnetic phase,    with no confinement and
no  spontaneous flavor symmetry breaking.   In these vacua the low energy degrees of freedom are 
weakly interacting non-Abelian dual quarks and gauge particles,   as well as some  monopoles of products of 
$U(1)$ groups.   In $SO(n_c)$ 
theories, the situation is   qualitatively    similar \cite{CKKM}  to   $USp(2n_c)$  cases;  however,   the effective  gauge group
and  the unbroken global  group in the vacua in free-magnetic phase   are  given by
$ SO({\tilde n}_c)=SO(2n_f-n_c+4)$ and   $USp(2n_f)$, respectively, in   these   theories.

\section{Analyses \label{sec:analysis}}

Our analyses  leading to these results consist of  several independent  steps \cite{CKM,CKKM}:
   
\smallskip

\noindent  {i)}  Semi-classical analysis,  yielding the number of the vacua, ${\cal N}$;

 \noindent  {ii)}  Determination of dynamical symmetry breaking pattern at  $\mu \gg \Lambda$
 fixed   and   
 $ m_i \to 0$;
 
\noindent   {iii)} Check of the   correct decoupling of the adjoint fields  in the  limit,
 $\mu \to \infty$,  with $m_i$ and $N=1$  scale factor   $ \Lambda_1 \equiv  \mu^{n_c \over  
 3n_c-n_f}        \Lambda^{2n_c- n_f  \over 3n_c-n_f   } $  (for
   $SU(n_c)$
 for instance)   fixed; 
   
\noindent   {iv)} Study of $N=1$  vacua  at   $m_i, \mu   \ll \Lambda$,   from the Seiberg-Witten curves, 
    through   the mass perturbation around CFT  singularities \cite{AD};   
 
\noindent   {v)}  Study of $N=1$  vacua  at  $ m_i, \mu   \ll \Lambda$,    by using the 
 low-energy effective action \cite{ArPlSei,APS2},       leading to a clear   microscopic picture  of infrared physics;
 
 \noindent  {vi)} Numerical study of  the  maximal singularities of the Seiberg-Witten curves
 in the cases of rank $2$  gauge groups ($SU(3)$, $USp(4)$, $SO(4)$, $SO(5)$);
 
\noindent   {vii)}   Study of monodromy  around   the singularities  of the curves;
 
\noindent   {viii)}     Semi-classical  analysis of  the monopole   flavor multiplet structure 
 \`a la Jackiw-Rebbi.  

\smallskip

  These independent analyses lead,  in a remarkably subtle way,   to  mutually  consistent answers as regards
the number of the vacua and     physical properties of each of them.    
We shall  discuss  the two among  the most important   aspects    of  our analysis,    ii) and v), 
below, 
but  let us   mention here   the results of  the semiclassical study, i).  As the models have no flat
directions, one can simply minimize  the classical potential and get, after taking into account the 
appropriate    Witten's
index in case part of the gauge group remains unbroken by classical VEVS,  the number of vacua.  For
$SU(n_c)$ theory with $n_f$ flavors,  there are    
\begin{equation}
{\cal N} = \sum_{r=0}^{{\hbox {\rm min}} \, \{n_f, n_c-1\}}\, (n_c-r)
\, \pmatrix{n_f \cr r}
\label{nofvac}
\end{equation}
classical  solutions.
Note that when $n_f$ is equal to or less than $n_c$ the sum over $r$
is done readily, and Eq.~(\ref{nofvac}) is equal  to
\begin{equation}
{\cal N}_1= ( 2 \, n_c - n_f) \, 2^{n_f -1}, \qquad (n_f \le n_c) \, .
\label{nofvacbis}
\end{equation}
Similarly,  we find  for $USp(2n_c)$
\begin{equation}
{\cal N} = \sum_{r=0}^{min\{ n_c, \, n_f\}}  (n_c- r +1) \cdot
\pmatrix {n_f \cr  r}. \, 
\label{Nvspnclass}   
\end{equation} 
vacua,    which reduces 
for smaller values of $n_f$ to  a closed expression, 
\begin{equation} {\cal N} =(2 \, n_c+2-n_f) \, 2^{n_f-1},  \qquad   (n_f \le n_c).
\label{Nvspnclassbis} 
\end{equation}   
For $SO(n_c)$ theories the result is \cite{CKKM}  
 \beq {\cal N} =    \sum_{r=0}^{min\{[n_{c}/2], n_{f}\}}
  w(n_{c}-2r)  \pmatrix{n_{f} \cr   r}   +     {n_f \choose n_c/2 },
  \label{sontotal}\eeq
 where 
 \beq   w(N)= N-2, \quad   N\ge  5,  \eeq
  and  $w(N)=4,\, 2,\, 1,\, 1,\, 1,\,$ for  $N= 4,\, 3,\, 2,\, 
  1,\,0,\,$ respectively,  and the last term is present only for  
$  2n_f \ge n_c,   \,\,\,    n_c= {\hbox{\rm  even}}. $
  The formulas Eq.(\ref{nofvac})-Eq.(\ref{Nvspnclassbis}) 
correctly reduce to   the well-known result
 \begin{equation}
    {\cal N} = n_f +2, \label{nvacsu2}
\end{equation}
in the case of the   $SU(2)$  theory.    Note   that the generalization is
nontrivial: e.g., $ {\cal N}
\ne n_f +n_c$  in $SU(n_c)$ theory!      The complexity of the formulae  Eq.(\ref{nofvac}) -
 Eq.(\ref{sontotal}) as compared
to Eq.(\ref{nvacsu2})      signals   indeed the presence of a rich variety  of  dynamical possibilities   in  general $SU(n_c)$,
$USp(2n_c),$  and $SO(n_c)$   theories,   some of    which might well be important in the understanding of the
standard QCD.

\section{Determination of dynamical symmetry breaking pattern at large $\mu$} 

At large  $\mu$  ($\mu \gg \Lambda$),
 the effective superpotential    can be read off from the
bare Lagrangian by integrating out the heavy, adjoint fields and by
adding to  it   the known exact instanton--induced superpotentials of the
corresponding $N=1$ theories.     
$N=1$ supersymmetry  guarantees that  physics depend on $\mu$ holomorphically,
so there cannot be any phase transition   as $\mu$ is varied to smaller values.

\subsection{ $SU(n_c)$:  $n_f \le  n_c+1$} 

When the number of flavors is relatively small, the effective superpotential
takes the form: \footnote{To be precise, this is the form of the superpotential for generic $n_f < n_c$. For
special cases $n_f = n_c$ and   $n_f= n_c +1,$  one must use appropriate superpotentials involving 
baryonlike  composites as well as mesons.  See \cite{CKM}. }
\begin{equation}
W =  -{1 \over 2 \mu} \left[ \Tr M^2 - {1 \over n_c}(\Tr M)^2 \right]  + \Tr (M m )  +
(n_c-n_f)  {\Lambda_1^{(3n_c - n_f)/(n_c-n_f)}  \over (\det M)^{1/(n_c-n_f)}},   
\label{splarmusmnf}  
\end{equation}
where  $M_j^i  \equiv   {\tilde Q}_j Q^i $  are the $n_f \times n_f$  meson matrix, and   
\beq  \Lambda_1 \equiv  \mu^{n_c \o   3n_c-n_f}        \Lambda^{2n_c-
n_f  \o 3n_c-n_f   } \eeq
is the invariant mass scale  of the $N=1$  SQCD (without the adjoint fields). 
The last term of Eq.(\ref{splarmusmnf})  is the Affleck-Dine-Seiberg instanton-induced 
superpotential,  the first arises   from integrating out the adjoint  fields $\Phi$,
$m= \diag(m_1, m_2, \ldots, m_f)$  is the bare quark mass matrix.  After making an Ansatz, 
$\bra M\ket =  \diag (\lambda_1, \lambda_2, \ldots, \lambda_{n_f}),$  one can straightforwardly  find the
minima of the potential.  We find   $ \,( 2 n_c- n_f )  {}_{n_f}\!C_{r}  $  vacua  in which 
the global symmetry is spontaneously broken (in  $m_i \to 0$ limit)  as
\begin{equation}
U(n_f) \to U(r) \times U(n_f - r).
\end{equation}
   By summing over $r$, $r=0,1,2,\ldots  [{n_f \o 2}]$,   one finds  ${\cal N}_1$ of  
Eq.(\ref{nofvacbis}).

\subsection {$SU(n_c)$:  $n_f \ge  n_c+2$}

When the number of flavor exceeds  $n_c +1$,   physics at low energies   is described by the 
effective  superpotential, 
\beq    {\cal W} = {\tilde q} M q  +    \Tr ( m M )   -{1 \o 2 \mu} \left[ \Tr
M^2 - {1 \o n_c}(\Tr M)^2 \right],
\label{dualqeq}\eeq
where  $q$ stands for the $n_f$   dual quarks in the fundamental representation of  the dual
gauge group  $SU({\tilde n}_c)$, $M$ is   the meson matrix as in   Eq.(\ref{splarmusmnf}). 
By first  minimizing the potential with respect to   $q$ and $M$, one finds    
\beq
{\cal N}_2= \sum_{r=0}^{{\tilde n}_c-1} \,
_{n_f}\!  C_ r \, ( {\tilde n}_c-r )
\label{extrav}
\eeq
of vacua,  in which    VEVS  behave as 
\begin{equation}
\bra q \ket \to 0,\qquad   \bra M  \ket \to 0, 
\end{equation}
in the limit, $m_i \to 0$.   
In other words, 
in these  vacua the global $SU(n_f)\times U(1)$  symmetry remains unbroken.

One seems to encounter a puzzle though:    the number of the vacua found here  ${\cal N}_2$
is always  less than  the known total number of vacua, ${\cal N}$  (Eq.(\ref{nofvac})).  Where 
are other vacua?

Actually,    we have tacitly assumed     $ \rank \, M <  n_f$ above, for otherwise
the dual quarks are all massive and the  theory reduces to the  pure $SU({\tilde n}_c)$ Yang-Mills in the 
infrared: its strong interaction  dynamics must be taken into account in order to get
information about its ground state.\footnote{ In fact,  a related puzzle is   how Seiberg's dual
Lagrangian
\cite{Sei} - the first two terms of  Eq.~(\ref{dualqeq})  -  can   give    the
right number of vacua for  the   massive
$N=1$   SQCD  with  $n_f > n_c+1$.        By following the same method as below
but with   $\mu = \infty$,
   we do   find the correct
number ($n_c$)    of vacua.  }   In order to retrieve these vacua, we must     first integrate   out the
dual quark fields. The instanton effects  in the dual gauge group $SU({\tilde  n}_c)$  leads   to  a 
superpotential, which is identical  to  (actually continuation of ) 
$W$  in Eq.(\ref{splarmusmnf})!     The minimization of such a potential yields    $ \,
{\cal N}_1 = ( 2 \, n_c - n_f) \, 2^{n_f -1}
\,  $       vacua  as before.

But then, for consistency,   the sum of  ${\cal N}_1$ and   ${\cal N}_2$  must be equal to  ${\cal N}$ 
of   Eq.(\ref{nofvac}).     As one can show easily by changing the dummy  variable  and  by using the known
identities among the binomial coefficients,   this    is indeed so.

To sum up, 
 in   $SU(n_c)$
theories  the 
exact global $U(n_f)$ symmetry in the equal mass  (or massless)
limit, which is spontaneously broken
to $   U(r)\times
U(n_f-r)  $     in  $ \, (2n_c -n_f)\,  {}_{n_f}\!C_ r\, $ vacua,
$r=0,1,\ldots, [n_f/2]$.
When the number of the flavor  is larger ($n_f > n_c +1$),     we find another class  of vacua,  with   no  global symmetry
breaking.   We shall see that these match the vacua in the free-magnetic phase  at small $\mu$.

\subsection{$USp(2n_c)$, $SO(n_c)$ }

The analysis in the cases of $USp(2n_c)$ or $SO(n_c)$   theories is similar, although the results are 
qualitatively different from the case of  $SU(n_c)$ theory.  

In $USp(2n_c)$   (or $SO(n_c)$)    theories, for small numbers of flavors, {\it the chiral
$SO(2n_f)$     (or $USp(2n_f)$) symmetry   in the massless limit is always spontaneously
broken     to   $U(n_f)$. }    This result nicely agrees with
what is expected generally from bi-fermion condensate of the standard
form in non supersymmetric theories, and forms a result in closest analogy with 
what is supposed to occur in the standard  QCD with small number of flavors.   

Finally,  in the cases of $USp(2n_c)$ or $SO(n_c)$   theories too,   there exist  also  other vacua without   dynamical symmetry
breaking,  when  the number of flavor is greater  ($n_f >  n_c+2$ or   $2 n_f > n_c-4$, respectively).

\section{Quantum vacua  at   $\mu \ll \Lambda$ }
  
At small $\mu$,  the infrared properties of the theory are described 
by  certain singularities of Seiberg-Witten  curves \cite{SW1,SW2,SUN}.   To be concrete take the  case of $SU(n_c)$
gauge  theory with $n_f$ flavors.    $N=1$  supersymmetric vacua are found by requiring that the curve
\begin{equation}
    y^{2} = \prod_{k=1}^{n_{c}}(x-\phi_{k})^{2} + 4 \Lambda^{2n_{c}-n_{f}}
    \prod_{j=1}^{n_{f}}(x+m_{j}),
\label{curve1} \end{equation}
where   \begin{equation}
\bra \phi \ket =\diag(\phi_1, \phi_2, \ldots), 
\end{equation}
describe the gauge invariant VEVS, 
\begin{equation}
u= \bra \Tr \Phi^2  \ket   =   \bra  \sum_{i<j} \phi_i \phi_j  \ket,  \quad  u_3= 
\bra \Tr \Phi^3  \ket   = \bra \sum_{i<j<k }
\phi_i \phi_j \phi_k  \ket, 
\end{equation}
etc.,   
to be maximally singular.   Since there are   $n_c-1$ free parameters,   up to  $n_c-1$
pairs of branch  points can be made  to coincide by appropriate choices of  $\{ \phi\}$, and   this
corresponds to the condition that there are maximal number of   massless monopoles of 
Abelian subgroup  $U(1)^{n_c-1}  \subset SU(n_c).$

Such a connection follows from the by now well understood association of monopole masses
with  the integral over canonical cycles of meromorphic differentials on the curve such as 
   Eq.(\ref{curve1}) \cite{SW1,SW2,SUN}.
Also one can show that, upon  $\mu$ perturbation, only these singular points of QMS (quantum space of
vacua)  lead to supersymmetric  ground states.

Once the low-energy degrees of freedoms (monopoles) are identified and   
their quantum numbers known,  it is in principle straightforward to analyze the 
properties of the vacua. 

It turns out that   the limit $m_i \to 0$   is highly   nontrivial, and  physics in the infrared 
  is far   richer
than  one might have expected  from the knowledge of $SU(2)$ gauge theory \cite{SW2}.   In particular,  one finds
that  the low-energy  degrees of  freedom are not always the  monopoles of the maximal Abelian 
subgroup envisaged in the  Nambu-'t Hooft-Mandelstam mechanism.

The crucial steps of our analysis are   the points   iv) and   v)  
of Sec.~\ref{sec:analysis}.
The first  of these steps  shows how all  the $N=1$ vacua, selected out
by the adjoint mass perturbation, 
 are associated  with  the
various universality classes of SCFT  \cite{AD},  and   allows us to relate  the quantum vacua  at small
$\mu$  to those at large
$\mu$;       the second step leads to  the  microscopic picture of confinement and dynamical symmetry breaking,  summarized
below.

\section{Microscopic Picture of Dynamical Symmetry Breaking}

  In $   SU(n_{c})$ theories with  $n_f$ flavors,  there are two group of  vacua. 
In the first group of vacua  with finite meson or dual quark vacuum
expectation values   (VEVS), labeled by an integer $r$, $r
\le [n_f/2]$,   the system is in confinement phase.   The nature of
the   actual carrier of   the flavor quantum numbers
depends on
$r$.   In  vacua with $r=0$,     magnetic
monopoles  are singlets of  the global $U(n_f)$  group,  hence   no
global symmetry breaking accompanies confinement.

In vacua with $r=1$,   the light particles are  magnetic  monopoles
in  the fundamental representation of $U(n_{f})$ flavor group.      Their condensation leads
to    the confinement and
flavor symmetry breaking,  simultaneously.

In vacua   labeled by $r$,  $2\leq r < n_{f}/2$
($r \neq n_{f} - n_{c}$),     the grouping of the associated
singularities  on the Coulomb
branch, with multiplicity, ${}_{n_f}\!C_r$,     at first
sight      suggests the condensation of monopoles in the rank-$r$
anti-symmetric tensor representation of the global
$SU(n_f)$ group.    Actually,   this does not occur.    The 
low-energy degrees of freedom of   these theories are
$n_f$   magnetic quarks  (in ${\underline r}$) 
   plus a  number of singlet monopoles  of  a non-Abelian   effective 
$SU(r)\times  U(1)^{n_c-r}$ gauge theory \cite{ArPlSei}.   

Monopoles in  higher representations of $SU(n_f)$ flavor group,  
even if they  exist
semi-classically,     break up into magnetic quarks    before they become
massless at  singularities on the Coulomb branch.     It is    the
condensation   of
the  latter   that    induces    confinement and  flavor
symmetry breaking,
$U(n_f) \rightarrow U(r) \times U(n_f-r)$,    in these vacua.   The system   thus   realizes  the  global  symmetry
of the theory in a Nambu-Goldstone mode,
       without
having      unusually    many Nambu-Goldstone
bosons.    It  is a novel   mechanism   for confinement  and
dynamical symmetry breaking.

   In the special cases with
$r= n_f/2$,   still  another  dynamical scenario takes place.   In these
cases,   the interactions among the monopoles and dyons  become  so strong
that  the low-energy theory describing them is a nontrivial SCFT,  with  conformal invariance explicitly broken by the adjoint
or quark masses.     Although the symmetry breaking pattern is known $U(n_{f})
\rightarrow U(n_{f}/2) \times U(n_{f}/2)$, the low-energy degrees of
freedom  in general    involve relatively nonlocal   fields and   their
interactions cannot be described  in terms of    a local
action.

Finally, there are    vacua in which
magnetic-quarks  do not  condense  and remain as  physically
observable particles at long distances, interacting with non-Abelian dual gauge particles
  (free magnetic phase),
when  the number of the flavor $n_f$  exceeds $n_c+1$.
The   global  $U(n_{f})$ symmetry  remains    unbroken in these vacua.  These precisely match  the 
second group of vacua found at large $\mu$ in which no condensate forms.

In $USp(2n_c)$ theories, again,  we find two groups of vacua,  whose  properties are shown in Table
\ref{tabuspn}.    The most salient  difference as compared to the $SU(n_c)$ theory   
is that here   the entire first group of vacua    corresponds to  
  a  SCFT.   It is a nontrivial superconformal theory:  one does 
not have a local effective Lagrangian description  for those 
theories.\footnote{Except for  $n_f=3$ and $n_f=2$, in which cases
we expect 
a local description to be  valid, see Sec. 8.2 of Carlino et. al. \cite{CKM}.   }    Nonetheless,   the symmetry
breaking pattern can be deduced,    from the analysis
 done at large $\mu$:    
   $SO(2n_f) $  symmetry is always  spontaneously  to   $U(n_f)$.

To see better what is going on, it is  instructive  to consider the  equal but nonvanishing   quark    mass
case  first.    
 See Table \ref{tabspnnzm}.     The flavor symmetry 
group of the underlying theory is now  explicitly broken  to $U(n_f)$.      The  first group of vacua 
 split   into various
subgroups of vacua  labeled by
$r$,   $r=0,1,2,\ldots, [{n_f -1\o 2}]$,    each of which   is described   by a {\it local}
  effective gauge theory 
of    Argyres-Plesser-Seiberg \cite{ArPlSei}  for $SU(n_c)$ theory (!),     with     gauge group  
$SU(r)
\times U(1)^{n_c -r +1 }   $   and $n_f$ (dual)  quarks in the  fundamental representation of   
$SU(r)$.\footnote{The key fact that some of the SCFT in
$USp(2n_c)$ or in $SO(n_c)$  theories at $m_i=m\ne 0$    are in the same universality classes
as those occurring in  the $SU(n_c)$ theory,  was  first noted by  Eguchi et. al. \cite{AD}.  Our perturbation theory in
masses around the SCFT and the effective action analysis  show that these SCFT   are indeed the $N=1$
vacua    which survive the adjoint mass perturbation (missed in the analysis of  \cite{APS2}) .  } 
Indeed, the gauge invariant composite VEVS  characterizing these  vacua   differ by some positive powers of $m$, so that 
the validity of each effective theory is limited  to small fluctuations  of order of $m^a,\,  $     $  a>0,\,$ around each
vacuum.

In the limit $m   \to 0$   these points  in the quantum moduli space (QMS)     collapse
into one single point    and accordingly    the range of the validity of each  local effective action shrinks to zero, implying 
that we have   a
nontrivial SCFT here,    with mutually nonlocal massless states.   
   In the example of $USp(4)$ theory with $n_f=4$, we have
explicitly verified   this by determining the singularities and branch
points at finite equal mass  $m$ and then    by studying    the limit $m
\to 0$.

\begin{table}[h]
\begin{center}
\vskip .3cm
\begin{tabular}{|l|c|c|c|}  
\hline  
    Deg.Freed.      &  Eff. Gauge  Group
&   Phase    &   Global Symmetry     \\
\hline
\hline
   monopoles   &   $U(1)^{ n_c} $               &   Confinement
   &      $U(n_f) $            \\ \hline
 monopoles         &  $U(1)^{n_c } $           &
Confinement       &     $U(n_f-1) \times U(1) $        \\ \hline
dual quarks  &   $SU(r)
\times U(1)^{n_c -r +1}      $  &    Confinement
&          $U(n_f-r) \times U(r) $  \\ \hline
dual quarks  & $ SU({n_f\over 2}) \times U(1)^{n_c -{n_f\over 2} +1}$   & SCFT & $U({n_f\over 2}) \times U({n_f\over 2}) $
   \\ \hline
\end{tabular}
\caption{ The first group of vacua  of $USp(2  n_c)$    theory with $n_f$ flavors  
 with $m_i=m \ne 0$.  The vacuum label $r$ in the third row runs for $r= 2,3,\ldots, [{n_f -1 \over 2}]$.    }
\label{tabspnnzm}
\end{center}
\end{table}

The first group of vacua  in the $SO(n_c)$  gauge theory has similar characteristics as the one 
in $USp(2n_c)$ theory   just discussed.      These  cases, 
together with the special
$r=n_f/2$  vacua  for the $SU(n_c)$ theory,   reveal   
another   new mechanism for dynamical symmetry breaking:   although
the global symmetry breaking
pattern  deduced  indirectly  looks familiar enough,  the low-energy
degrees of freedom are  relatively nonlocal dual
quarks and dyons.  It would be interesting to get a better
understanding   of    this phenomenon.

    For large
numbers of flavor,  there are  also     vacua,  just as in large
$n_f$  $SU(n_c)$  theories, with no confinement and no  dynamical
flavor symmetry breaking.   Physics around  these vacua can be explicitly studied by
use of the effective low-energy action  at the "special points"   of  \cite{APS2}.   
The low-energy particles are  solitonlike
magnetic  quarks which weakly interact  with  dual (in
general) non-Abelian gauge fields: the system  is in the free magnetic phase.   These 
are smoothly connected to the second group of vacua, with no symmetry breaking,   
found at large $\mu$.   

\begin{figure}[h]
\begin{center}
\epsfxsize=24pc 
\epsfbox{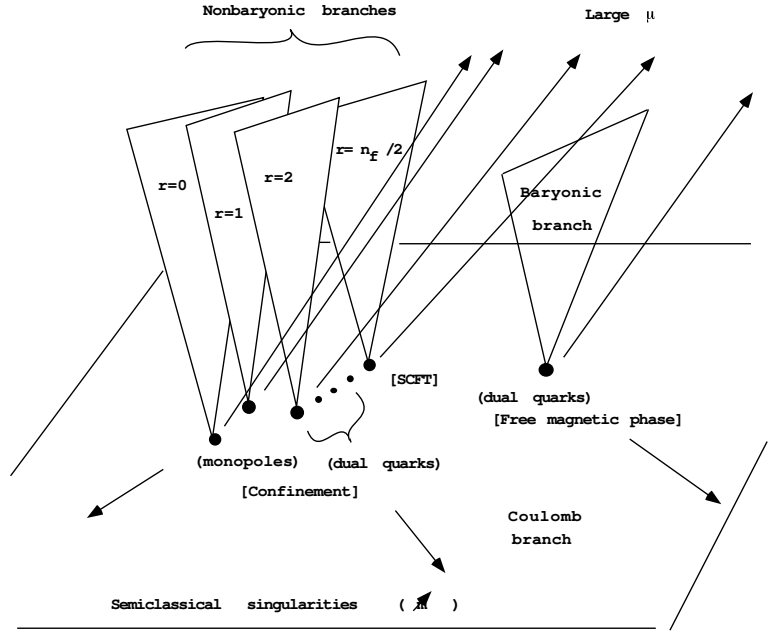}   
\caption{A  schematic view of  the $N=2$   space of vacua  (QMS) in  $SU(n_c)$  theories    and  
its   singularities corresponding to
$N=1$ vacua    (black dots).   The latter lie   near  the  roots of various Higgs branches.  
When the generic quark masses are added  each  of $N=1$ vacuum   further splits. At large quark masses 
these points 
move to semiclassical  regions of Coulomb branch.  }  
\label{fig:sun}
\end{center}
\end{figure}

\begin{figure}[h]
\begin{center}
\epsfxsize=24pc 
\epsfbox{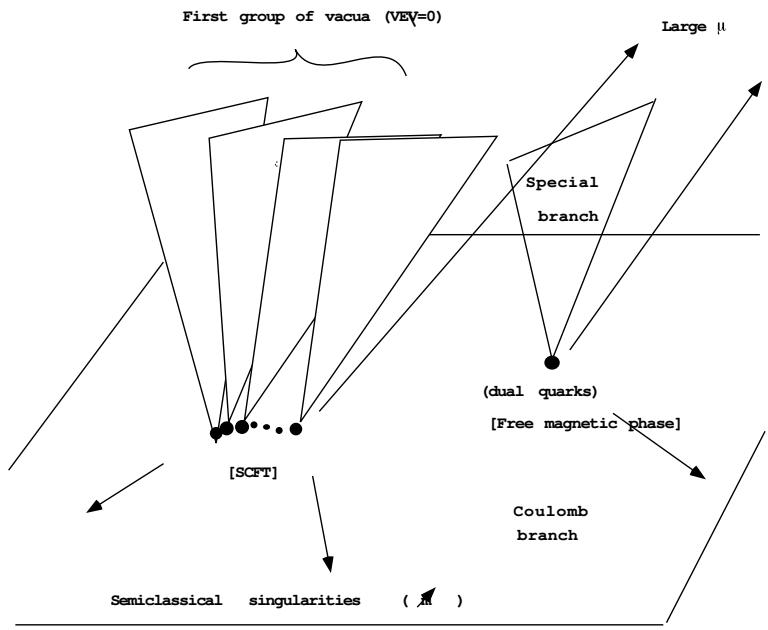}       
\caption{$N=2$  space of vacua  (QMS)  and   the singularities corresponding to $N=1$ vacua   
(black dots) near  the  roots of various Higgs branches,  in $USp(2n_c)$  theories.   For $m_i=m\ne 0$
the situation is similar to $SU(n_c)$ case;   in the $m \to 0$ limit, however, the various vacua
of the first group collapse into one single vacuum,  leading to a nontrivial  fixed point behavior. 
\label{fig:usp}}
\end{center}
\end{figure}

Properties of  various vacua  in  $SU(n_c)$  and $USp(2n_c)$   theories are 
illustrated    schematically in   Fig.~\ref{fig:sun} and  Fig.~\ref{fig:usp}.

\section*{Acknowledgments}
The author thanks  G. Carlino,  P. Kumar and H. Murayama  for fruitful and enjoyable 
collaborations, and  
 the organizers of the workshop ``Continuous Advances in QCD" (Univ. Minnesota,  May 2000)
 for inviting him  to present and discuss  these  recent results  in  such a stimulating  atmosphere.

\bigskip

\bigskip

\bigskip

\bigskip

\end{document}